\documentclass[pra,aps,showpacs,twocolumn,floatfix, superscriptaddress, nofootinbib, nobibnotes]{revtex4-1}
\usepackage{graphicx}
\usepackage[ansinew]{inputenc}
\usepackage{array}
\usepackage{color}
\usepackage{amsmath}
\usepackage{amsxtra}
\usepackage{amstext}
\usepackage{amssymb}
\usepackage{latexsym}
\usepackage{dsfont}

\begin{document}

\title{Quantum reservoir engineering through quadratic optomechanical interaction in the reversed dissipation regime}

\author{Jae Hoon Lee}
\affiliation{Center for Time and Frequency, Division of Physical Metrology, Korea Research Institute of Standards and Science, Daejeon, 34113, South Korea}
\author{H.~Seok}
\email{hseok@kongju.ac.kr}
\affiliation{Department of Physics Education, Kongju National University, Gongju 32588, South Korea}

\begin{abstract}
We explore the electromagnetic field coupled to a mechanical resonator via quadratic optomechanical interaction in the reversed dissipation regime where the mechanical damping rate is much larger than the cavity field dissipation rate. It is shown that in this regime, the cavity field effectively acquires an additional reservoir which is conditioned by the temperature of the mechanical bath as well as the mechanical damping rate. We analytically find the steady-state mean photon number and the critical temperature of the mechanical oscillator to cool or heat the coupled electromagnetic field. We also show that in the case of quadratic coupling, the temperature of the mechanical oscillator can be estimated in the quantum regime by observing the noise spectrum of the cavity field.
\end{abstract}

\maketitle

\section{Introduction}
Engineering a quantum mechanical system, for which the interaction between light and the motional degrees of freedom of a mechanical oscillator can be extensively controlled, is the objective of the fast developing research field of cavity optomechanics ~\cite{ANDP:ANDP201200226, RevModPhys.86.1391, Milburn_optomechanics_book}. Various types of cavity optomechanical devices have been developed to take advantage of this interaction for applications in precision measurement of mechanical motion~\cite{Purdy801,  Schreppler1486}, exploration of macroscopic objects in the quantum regime~\cite{PhysRevA.84.052121}, and fundamental platform research of hybrid quantum systems~\cite{PhysRevLett.107.223001, PhysRevLett.112.133603}.
Recently, rapid advancements in fabrication methods for various mechanical elements have led to the development of high quality factor micro- and nano-scaled devices for a broad range of parameters~\cite{ANDP:ANDP201200226, RevModPhys.86.1391}. With these on-chip devices, it is now possible to cool a mechanical oscillator to its motional ground state~\cite{Teufel:2011aa, Chan:2011aa}, exhibit optomechanically induced transparency~\cite{PhysRevA.81.041803, Weis1520}, coherently couple optical and mechanical modes~\cite{Verhagen:2012aa}, produce entanglement between optical and mechanical resonators~\cite{Palomaki710}, optically mediate interaction between mechanical oscillators~\cite{PhysRevLett.112.013602, Spethmann:2016aa}, generate squeezed light~\cite{Brooks:2012aa, Safavi-Naeini:2013aa, PhysRevX.3.031012}, and perform precision measurement of the mechanical motion near the standard quantum limit~\cite{Schreppler1486}.

Various new phenomena observed in quantum optomechanical systems are being evaluated from the point of view of reservoir engineering~\cite{PhysRevLett.77.4728, PhysRevLett.107.080503, Kienzler53}. 
Most optomechanical experiments to date are in the so-called normal dissipation regime where the optical energy dissipation rate is much larger than the mechanical energy decay rate. 
In this regime, the mechanical oscillator is effectively coupled to the optical reservoir via optomechanical interaction. The coupling strength and the effective temperature of the optical reservoir seen by the mechanics can be engineered by controlling the optomechanical interaction. More recently, the notion of a reversed dissipation regime has been introduced, which refers to a regime where the mechanical energy dissipation rate is much larger than that of the cavity field energy~\cite{PhysRevLett.113.023604}. Particularly, mechanical reservoir engineering in this regime via linear optomechanical coupling has been demonstrated showing its ability to tailor microwave fields for cooling or amplification~\cite{Toth:2017aa}.

In this paper, we theoretically analyze an optomechanical system in which an electromagnetic field and a mechanical oscillator are coupled via quadratic interaction~\cite{PhysRevA.82.021806, Sankey:2010aa, PhysRevX.5.041024}. 
In particular, a single-mode of the mechanical oscillator is quadratically coupled to a single-mode cavity field in the reversed dissipation regime. This results in an additional effective reservoir for the cavity field after adiabatic elimination of the mechanics. Calculations show that the coupling strength between the cavity field and the effective reservoir is dependent on the temperature of the mechanics. An analytic solution for the critical temperature of the mechanical oscillator is presented indicating the threshold temperature for which the electromagnetic field is cooled. Linear and quadratic interactions in optomechanical systems,  which produce distinct cavity field noise spectra, are comparatively investigated.

The remainder of this paper is organized as follows:
The model system is described in Sec.~\ref{sec:Model System} and the effective master equation for the cavity field is derived in the reversed dissipation regime in Sec.~\ref{sec:Formalism}. In Sec.~\ref{sec:Results}, we discuss the broadening of the linewidth for the cavity field and analytically find the steady-state mean photon number of the cavity field for both linear and quadratic optomechanical systems. Finally, Sec.~\ref{Sec:Conclusions} gives our summary and conclusions. 

\section{Model System}
\label{sec:Model System}
We consider an optomechanical system in which the single mode of a mechanical resonator of effective mass $m$ and frequency $\omega_m$ is quadratically coupled to a single-mode cavity field of frequency $\omega_c$, inside an optical resonator driven by a monochromatic field of frequency $\omega_{L}$ at pumping rate $\eta$. In the frame of the pump frequency for the cavity field, the system Hamiltonian describing the optomechanical system is given by 
\begin{eqnarray}
 \hat H_{\rm sys} &=& -\hbar\Delta_c \hat a^\dag \hat a + i\hbar(\eta\hat a^\dag- \eta^*\hat a) \nonumber  \\ 
 &&+ \hbar\omega_m \hat b^\dag\hat b+\hbar g_0^{(2)}\hat a^\dag \hat a (\hat b+\hat b^\dag)^2,
\end{eqnarray}
where the first two terms describe the driven single-mode optical field, the third term describes the Hamiltonian for the free mechanical mode, and the last term depicts the quadratic optomechanical interaction. Here, $\Delta_c=\omega_L-\omega_c$ is the pump detuning from the cavity resonance, $g_0^{(2)}$ is the quadratic single-photon optomechanical coupling coefficient, and $\hat a$ ($\hat b$) is the annihilation operator for the optical (mechanical) resonator.    

In order to properly describe an open optomechanical system in the quantum regime, the interaction between the optomechanical system and a coupled reservoir should be taken into account. This additional interaction with a reservoir accounts for photon loss and mechanical thermal fluctuations in the weak coupling regime. In the Schr{\"o}dinger picture, the dynamics of the open optomechanical system can be described by the master equation~\cite{Pierre_book}
\begin{equation}
\frac{d\tilde\rho}{dt} =({\cal L}_{\rm sys}+{\cal L}_o +{\cal L}_m)\tilde\rho, 
\end{equation}
where  $\tilde\rho$ is the density operator for the optomechanical system, the unitary evolution of the system is governed by the Liouvillian superoperator
\begin{equation}
 {\cal L}_{\rm sys} = -\frac{i}{\hbar}[\hat H_{\rm sys},~\cdot~]
\end{equation}
and the incoherent evolutions for the optical and mechanical fields are described by the sum of the standard Lindblad superoperators 
\begin{eqnarray}
{\cal L}_o &=&\frac{\kappa}{2}(\bar n_o+1)\{2(\hat a \cdot \hat a^{\dag})-(\hat a^\dag \hat a  \cdot)- (\cdot\hat a^\dag \hat a)\} \nonumber \\
&&+\frac{\kappa}{2}\bar n_o\{2(\hat a^{\dag} \cdot \hat a)-(\hat a \hat a^\dag  \cdot)- (\cdot\hat a \hat a^\dag)\},
\end{eqnarray}
\begin{eqnarray}
{\cal L}_m &=&\frac{\gamma}{2}(\bar n_m+1)\{2(\hat b \cdot \hat b^{\dag})-(\hat b^\dag \hat b  \cdot)- (\cdot\hat b^\dag \hat b)\} \nonumber \\
&&+\frac{\gamma}{2}\bar n_m\{2(\hat b^{\dag} \cdot \hat b)-(\hat b \hat b^\dag  \cdot)- (\cdot\hat b \hat b^\dag)\},
\end{eqnarray}
respectively. Note that the superoperators act on the density operator and the dot in the superoperators indicate where the density operator is to be placed. For typographical simplicity, it is convenient to make use of the dot notation in the superoperators for obtaining the solutions to the master equation and associated equations of motion~\cite{Carmichael_book_1}. ${\cal L}_o$ and ${\cal L}_m$ characterize the interaction of the cavity field and the mechanical oscillator with their associated reservoirs by accounting for absorption and emission of quanta from the reservoirs. Here the cavity field is damped by the optical bath with decay rate $\kappa$, and the mechanical oscillator is dissipated to the mechanical bath with decay rate $\gamma$. The thermal occupation numbers of the mechanical bath and the optical bath are denoted by ${\bar n}_m=[{\rm exp}(\hbar\omega_m/k_BT_o)-1]^{-1}$ and ${\bar n}_o=[{\rm exp}(\hbar\omega_c/k_BT_m)-1]^{-1}$, respectively. Here, $k_B$ is the Boltzmann constant while $T_o$ and $T_m$ are the temperatures of the optical and mechanical heat baths, respectively.

\section{Formalism}
\label{sec:Formalism}
This section discusses the dynamics of the optomechanical system under the influence of fluctuations and noise in the quantum regime, along with the effective dynamics of the cavity field in the reversed dissipation regime.

\subsection{Interaction picture in the weak coupling regime}
For a cavity field that is externally driven by a classical field, it is beneficial to introduce the unitary displacement operator 
\begin{equation}
\hat U_1 =e^{(\alpha\hat a^\dag -\alpha^* \hat a)},
\end{equation}
with steady-state cavity field amplitude
\begin{equation}
\alpha=\frac{\eta}{-i\Delta_c+\kappa/2}.
\label{eq:Lorentzian}
\end{equation}
The master equation for the density operator $\bar\rho =  \hat U_1^\dag   \tilde\rho\hat U_1$ in the displaced picture thus becomes
\begin{eqnarray}
\frac{d\bar \rho}{dt} =({\cal L}_{\rm sys}' +{\cal L}_o+{\cal L}_m)\bar\rho, 
\end{eqnarray}
where
\begin{equation}
\begin{aligned}
 {\cal L}_{\rm sys}' =   i\left[\Delta_c\hat a^\dag\hat a-\omega_m'\hat b^\dag\hat b+g_0^{(2)}|\alpha|^2(\hat b^{\dag 2}+\hat b^2) \right. \\
\left.  -g_0^{(2)}(\alpha^*\hat a+\alpha\hat a^\dag)(\hat b^{\dag}+\hat b)^2-g_0^{(2)}\hat a^\dag\hat a(\hat b^{\dag}+\hat b)^2,~\cdot~\right],
\end{aligned}
\end{equation}
where $\omega_m^\prime = \omega_m + 2g_0^{(2)}|\alpha|^2$ is the shifted frequency of the mechanical oscillator. The static mechanical frequency shift, which is proportional to the cavity photon number, results in a significant modification of the stiffness of the mechanical oscillator. Therefore we confine the single-photon quadratic optomechanical coupling coefficient $g_0^{(2)}$ to be positive in order to avoid mechanical instability as well as to retain a high mechanical quality factor. Throughout this paper we consider a stable mechanical oscillator with a high quality factor in a harmonic potential.  

In a physically relevant regime in which the single-photon optomechanical coupling coefficient between the optical and mechanical resonator is small, and the steady-state mean amplitude of the cavity field is large, $|\alpha|\gg1$, the effect due to the superoperator $[g_0^{(2)}\hat a^\dag\hat a(\hat b^{\dag}+\hat b)^2,~\cdot~]$ can be negligible when describing the dynamics of the density operator. This approximation allows us to safely linearize the optomechanical interaction with respect to the cavity field operators. Moreover, when exploring the optomechanical system in the quantum regime, we assume that the optomechanical system is in the resolved-sideband limit ($\omega_m'\gg \kappa$), and that the external driving field is red-detuned by twice the shifted mechanical frequency, $\Delta_c=-2\omega_m'$. This condition is fulfilled when
\begin{equation}
\Delta_c^3+2\omega_m\Delta_c^2+\frac{\kappa^2}{4}\Delta_c+\frac{\omega_m\kappa^2}{2}+4g_0^{(2)}|\eta|^2=0. 
\label{eq:resonance_condition}
\end{equation}

\begin{figure}[]
\includegraphics[width=0.48\textwidth]{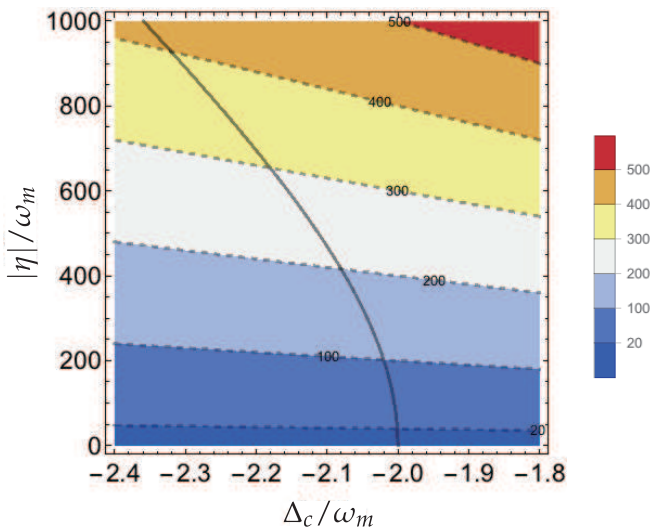}
\caption{Magnitude of the steady-state cavity field amplitude $|\alpha|$ as functions of cavity detuning $\Delta_c$ and pumping rate $|\eta|$ normalized with mechanical frequency $\omega_m$. The thick solid line indicates the resonance condition $\Delta_c=-2\omega_c$. The parameters we use are $g_0^{(2)}/\omega_m=5\times10^{-7}$, $\kappa/\omega_m=0.0025$. }
\label{fig:parameter}
\end{figure}

Figure~\ref{fig:parameter} shows the solution of Eq.~(\ref{eq:resonance_condition}) along with $|\alpha|$ as functions of cavity detuning $\Delta_c$ and pumping rate $|\eta|$ according to Eq.~(\ref{eq:Lorentzian}), for the given parameters $g_0^{(2)}/\omega_m = 5\times 10^{-7},~\kappa/\omega_m = 0.0025$. As expected, larger pumping rates increase the intracavity photon number which, in turn, gives large frequency shifts to the mechanical oscillator. Thus, the pump frequency would need to be further away from the cavity resonance so as to satisfy the resonance condition $\Delta_c=-2\omega_m'$. However, one should keep in mind that large detunings decrease the intracavity photon number due to the Lorentzian response of the cavity field amplitude, Eq.~(\ref{eq:Lorentzian}). Therefore, it is critical to find a valid parameter regime in which the resonance condition for the pump detuning, as well as the large cavity field requirement ($|\alpha|\gg1$) are simultaneously fulfilled.

Finally, by invoking the rotating-wave approximation in the interaction picture achieved by the unitary operator $\hat U_2 = e^{i(\Delta_c\hat a^\dag\hat a-\omega_m^\prime\hat b^\dag\hat b)t}$, we are able to significantly simplify the master equation as
\begin{equation}
\frac{d\rho}{dt} = ({\cal L}_{om}+{\cal L}_o+{\cal L}_m)\rho, \label{master}
\end{equation}
where $\rho = \hat U_2^{\dag}\bar\rho\hat U_2$ and 
\begin{equation}
 {\cal L}_{om} =- ig_2[\hat a^\dag\hat b^2+\hat b^{\dag 2}\hat a,~\cdot~].
\end{equation}
Here, we take $\alpha$ to be positive without loss of generality and $g_2=g_0^{(2)}\alpha$ is the field-amplified optomechanical coupling strength. It should be noted that we have neglected counter-rotating and off-resonant contributions in the regime where $g_0^{(2)}|\alpha|^2 \ll \omega_m'$. In other words, the rotating-wave approximation validates itself if the static frequency shift is much smaller than the bare mechanical frequency. Therefore, under the rotating-wave approximation, the cavity field can have large amplitudes in the weak-coupling and resolved-sideband regime.
Note that the superoperator ${\cal L}_{om}$ accounts for the mixing process in which one photon in the cavity field is annihilated (created) by the creation (destruction) of two phonons in the mechanical resonator. This process is equivalent to well-known corresponding mechanisms in a parametric amplifier in the quantum regime~\cite{Milburn_book_1}.

\subsection{Reduced density operator for the cavity field}
In order to find the effective master equation for the cavity field when the mechanical damping is the prevailing source of dissipation, we proceed to eliminate the mechanical variables, resulting in a master equation for the reduced density operator for the optical field. This technique is commonly used for eliminating the density operator for the pump mode of a parametric amplifier in quantum optics and for the optical field in cavity QED~\cite{Carmichael_book_1, Carmichael_book_2}. 

We first transform the density operator as
\begin{equation}
 \rho'(t) = e^{-({\cal L}_o+{\cal L}_m)t}\rho(t),
 \label{eq:density_transformation}
\end{equation}
and therefore the equation of motion of the transformed density operator reads
\begin{equation}
\frac{d\rho'}{dt} = {{\cal L}_{om}'}(t)\rho',
 \label{eq_master_starting}
\end{equation}
where ${{\cal L}_{om}'}(t) = e^{-({\cal L}_o+{\cal L}_m)t}{\cal L}_{om}e^{({\cal L}_o+{\cal L}_m)t}$. Owing to the fact that the two superoperators accounting for the dissipative evolutions for the cavity and mechanical resonators commute, $[{\cal L}_o, {\cal L}_m]=0$, ${\cal L}_{om}'(t)$ can be explicitly written as 
\begin{equation}
 {{\cal L}_{om}'}(t) = -ig_2\left(O_2(t)(\hat b^{2}\cdot)'+(\hat b^{\dag 2}\cdot)'O_1(t)-H.c.\right), 
 \label{eq:optomech_superoperator}
\end{equation}
where 
\begin{eqnarray}
 (\hat b^{\dag 2}\cdot)'  &=& e^{-{\cal L}_m t}(\hat b^{\dag 2}\cdot)e^{{\cal L}_m t}, 
 \label{eq:superoperator_1} \\
 (\hat b^{2}\cdot)' &=& e^{-{\cal L}_m t}(\hat b^{2}\cdot)e^{{\cal L}_m t}, \\
O_1(t) &=& (\hat a\cdot)' = e^{-{\cal L}_o t}(\hat a\cdot)e^{{\cal L}_o t}, \\
 O_2(t) &=&(\hat a^\dag\cdot)' = e^{-{\cal L}_o t}(\hat a^\dag\cdot)e^{{\cal L}_o t},
\end{eqnarray}
where we have factorized the superoperators, for example, $(\hat a\hat b^{\dag 2}\cdot)'=(\hat a\cdot)'(\hat b^{\dag 2}\cdot)'$. Notice that the primed superoperators, such as $ (\hat b^{\dag 2}\cdot)'$, are explicitly time-dependent while the unprimed superoperators, such as $(\hat b^{\dag 2}\cdot)$, are time-independent. 

Let us find the exact time-dependence of the superoperators involving the bosonic creation and annihilation operators for the mechanics. Taking the time derivative on both side of Eq.~(\ref{eq:superoperator_1}), we have 
\begin{equation}
 \frac{d}{dt}(\hat b^{\dag 2}\cdot)' = -{\cal L}_m(\hat b^{\dag 2}\cdot)'+(\hat b^{\dag 2}\cdot)'{\cal L}_m = \left[(\hat b^{\dag 2}\cdot)', {\cal L}_m'\right], 
\end{equation}
where in the second equality we have used the fact that a superoperator commutes with any function of it and thus ${\cal L}_m={\cal L}_m'$. In this fashion, one can build the closed set of differential equations for $\vec B(t) = \left( (\hat b^{\dag 2}\cdot)',  (\hat b^{\dag}\cdot\hat b^{\dag})',  (\cdot\hat b^{\dag 2})'\right)^T$ as 
\begin{equation}
 \frac{d}{dt}\vec B(t) = M\vec B(t),
 \label{eq:closed_set_of_eqns}
\end{equation}
where 
\begin{equation}
 M = 
 \begin{pmatrix}
  \gamma(2\bar n_m+1) & -2\gamma(\bar n_m+1) & 0  \\
  \gamma\bar n_m  & 0 & -\gamma(\bar n_m+1)  \\
  0  & 2\gamma\bar n_m  & -\gamma(2\bar n_m+1) 
   \end{pmatrix}.
\end{equation}
The solution to Eq.~(\ref{eq:closed_set_of_eqns}) can be obtained as 
\begin{equation}
 \vec B(t) = e^{M t}\vec B(0)\equiv S(t)\vec B(0), 
\end{equation}
where $\vec B(0) =  \left( (\hat b^{\dag 2}\cdot),  (\hat b^{\dag}\cdot\hat b^{\dag}),  (\cdot\hat b^{\dag 2})\right)^T$.  Noting that the $3\times3$ matrix $S(t)$ is a real, one can find the exact time-dependence of the superoperators in ${\cal L}_{om}'$ involving the bosonic creation and annihilation operators for the mechanics as 
\begin{flalign} \label{eq:time_dependence_1}
 (\hat b^{\dag 2}\cdot)' &= S_{11}(t)(\hat b^{\dag 2}\cdot)+S_{12}(t)(\hat b^{\dag}\cdot\hat b^{\dag})+S_{13}(t)(\cdot\hat b^{\dag 2}), \\
 (\hat b^{2}\cdot)' &= S_{31}(t)(\cdot\hat b^{2})+S_{32}(t)(\hat b\cdot\hat b)+S_{33}(t)(\hat b^{2}\cdot).
  \label{eq:time_dependence_2}
\end{flalign}
Substituting Eq.~(\ref{eq:time_dependence_1}) and~(\ref{eq:time_dependence_2}) into Eq.~(\ref{eq:optomech_superoperator}) gives rise to 
\begin{flalign}
  {{\cal L}_{om}'}(t) =& -ig_2\left(S_{11}(t)O_1(t)(\hat b^{\dag 2}\cdot)+S_{12}(t)O_1(t)(\hat b^{\dag}\cdot\hat b^{\dag})\right.\nonumber \\ &+S_{13}(t)O_1(t)(\cdot\hat b^{\dag 2})+ S_{31}(t)O_2(t)(\cdot\hat b^{2}) \nonumber \\
  &\left.+S_{32}(t)O_2(t)(\hat b\cdot\hat b)+S_{33}(t)O_2(t)(\hat b^{2}\cdot) -H.c.\right)
  \label{eq:26}
\end{flalign}
Notice in Eq.~(\ref{eq:26}), the mechanical superoperators do not have time-dependence, the time-dependence of the optomechanical coupling can be expressed with the optical superoperators and the matrix $S$ elements.

We are now in a position to formally solve Eq.~(\ref{eq_master_starting}) and to derive the effective master equation for the cavity field. The formal solution of Eq.~(\ref{eq_master_starting}) can be obtained as 
\begin{equation}
 \rho'(t) = \rho'(0) +\int_0^{t}d\tau  {{\cal L}_{om}'}(\tau)\rho'(\tau),
\end{equation}
and substituting the formal solution back into Eq.~(\ref{eq_master_starting}) results in
\begin{equation}
\frac{d\rho'}{dt} = {{\cal L}_{om}'}(t)\rho'(0) +\int_0^{t}d\tau  {{\cal L}_{om}'}(t){{\cal L}_{om}'}(\tau)\rho'(\tau).
\end{equation}
Taking a partial trace over the Hilbert space for the mechanics brings about the effective master equation for the reduced density operator for the optical field, 
\begin{equation}
\frac{d\rho'_o}{dt} = {\rm Tr}_m\{{{\cal L}_{om}'}(t)\rho'(0)\} +\int_0^{t}d\tau  {\rm Tr}_m\{{{\cal L}_{om}'}(t){{\cal L}_{om}'}(\tau)\rho'(\tau)\},
 \label{eq:reduce}
\end{equation}
where $\rho'_o = {\rm Tr}_m\{\rho'\}$ is the reduced density operator for the optical field and ${\rm Tr}_m\{\cdot\}$ denotes partial trace over the mechanical oscillator. 

In the regime where mechanical damping is the dominant source of dissipation, the state of the mechanics tends to approach thermal equilibrium in a timescale of $1/\gamma$. In this limit the density operator describing the optomechanical system can be approximated as a product state
\begin{equation}
\rho(t) \approx \rho_o(t) \otimes \rho_m(t),
\end{equation}
where $\hat \rho_o$ is the reduced density operator for the cavity field and $\hat \rho_m$ is the reduced density operator for the mechanics. On a timescale slower than $1/\gamma$, the dynamics of the optomechanical system can be solely represented by that of the cavity field, whereas the dynamics of the mechanical oscillator instantaneously follows that of the cavity field due to the fast mechanical dissipation. Specifically, the mechanical oscillator in thermal equilibrium can be described in the basis of Fock states as, 
\begin{equation}
 \rho_m =\sum_{n=0}^{\infty}\frac{\bar n_m^n}{(1+\bar n_m)^{n+1}}|n\rangle\langle n |, 
 \label{eq:mechanical_density}
\end{equation} 
where $| n\rangle$ are energy eigenstates of the free mechanical Hamiltonian. 
Noting that $\rho'(0)=\rho(0)$, one can find the first term in the right-hand-side of Eq.~(\ref{eq:reduce}) as
\begin{equation}
 {\rm Tr}_m\{{{\cal L}_{om}'}(t)\rho'(0)\}=0,
\end{equation}
due to the fact that all mechanical operators coupling to the cavity field have zero mean in the thermal state, i.e., ${\rm Tr}_m\{\hat b^{\dag2}\rho_m\}={\rm Tr}_m\{\hat b^2\rho_m\}=0$. However, the second term in the right-hand-side of Eq.~(\ref{eq:reduce}) has nonzero contribution and can be written as
  \begin{eqnarray}
 &&\int_0^{t}d\tau  {\rm Tr}_m\{{{\cal L}_{om}'}(t){{\cal L}_{om}'}(\tau)\rho'(\tau)\} \nonumber \\
 &&=g_2^2\int_0^{t}d\tau~e^{-\gamma(t-\tau)}\left\{2\bar n_m^2(O_2^\dag(t)O_2(\tau)-O_1(t)O_2(\tau))\right. \nonumber \\
 &&\left.+2(\bar n_m+1)^2(O_1^\dag(t)O_1(\tau)-O_2(t)O_1(\tau))\right\}+H.c.,
 \label{eq:second}
\end{eqnarray}
where we have used the nonzero expectation values ${\rm Tr}_m\{\hat b^{\dag2}\hat b^2\rho_m\}=2\bar n_m^2~, {\rm Tr}_m\{\hat b^{\dag}\hat b\rho_m\}=\bar n_m$. Concerning the dynamics of the optomechanical system on a timescale slower than $1/\gamma$, the exponential function in the time integral of Eq.~(\ref{eq:second}) decays very rapidly so that, in the adiabatic limit, it can be replaced with a $\delta$-function as
\begin{equation}
e^{-\gamma(t-\tau)} \rightarrow \frac{2}{\gamma}\delta(t-\tau).
\end{equation}
This limit is equivalent to the Markov approximation in that it eliminates memory effects on the optomechanical system. Now, the time integral can be performed analytically to give 
\begin{flalign}
\frac{d{\rho'_o}}{dt} =&\frac{4g_2^2}{\gamma}(\bar n_m+1)^2e^{-{\cal L}_o t}\{2(\hat a \cdot \hat a^{\dag})-(\hat a^\dag \hat a  \cdot)- (\cdot\hat a^\dag \hat a)\}\rho_o  \nonumber \\
&+\frac{4g_2^2}{\gamma}\bar n_m^2e^{-{\cal L}_o t}\{2(\hat a^{\dag} \cdot \hat a)-(\hat a \hat a^\dag  \cdot)- (\cdot\hat a \hat a^\dag)\}\rho_o,
\end{flalign}
where we have made use of the product rules for the superoperators, for example, $(\cdot\hat a)(\hat a^\dag\cdot)=(\hat a^\dag\cdot\hat a)$. Transforming the density operator back to the displaced frame, $\bar\rho_o=\hat U_2e^{{\cal L}_o t}\rho_o'\hat U^{\dag}$, we arrive at the effective master equation for the reduced density operator for the cavity field 
\begin{flalign}
\frac{d{\bar\rho_o}}{dt} =& i\Delta_c[\hat a^\dag\hat a,~ \bar\rho_o]+ \frac{D_e}{2}(2\hat a \bar\rho_o \hat a^{\dag}-\hat a^\dag \hat a  \bar\rho_o- \bar\rho_o\hat a^\dag \hat a)\nonumber \\
&+\frac{D_a}{2}(2\hat a^{\dag} \bar\rho_o \hat a-\hat a \hat a^\dag  \bar\rho_o- \bar\rho_o\hat a \hat a^\dag), 
 \label{eq:key_result}
\end{flalign}
where the emission coefficient $D_e$ and the absorption coefficient $D_a$ are given by
\begin{eqnarray}
 D_e &=& \kappa({\bar n_o}+1) + \frac{8g_2^2}{\gamma}({\bar n_m}+1)^2 , \\
D_a &=& \kappa{\bar n_o} + \frac{8g_2^2}{\gamma}{\bar n_m^2}. 
\end{eqnarray}
The derivation of Eq.~(\ref{eq:key_result}) which describes the dynamics of the optomechanical system in the reversed dissipation regime is the main result of this paper. Note that the cavity field in this expression is coupled to two independent reservoirs: the intrinsic optical  reservoir and an effective reservoir originating from the mechanical reservoir. In contrast to the dynamics in the normal dissipation regime where the optomechanical interaction provides an effective optical bath to the mechanical oscillator with coupling rate $g^2/\kappa$, 
Eq.~(\ref{eq:key_result}) shows that an effective mechanical reservoir is in contact with the cavity field with the coupling rate depending on $g^2/\gamma$ in the reversed dissipation regime.
In addition, it is by virtue of the quadratic optomechanical coupling that the coupling strength between the cavity field and the mechanical reservoir relies on the thermal occupation number of the mechanics. 

It is noteworthy to understand that the quadratic optomechanical interaction in the rotating-wave approximation, $H_{om}=\hbar g_2(\hat a\hat b^{\dag 2} +\hat b^2 \hat a^{\dag})$, is quadratic with respect to the mechanical operators while being linear in terms of the cavity operators. This discrepancy leads to a difference in how the system couples to its effective reservoir in normal and reversed dissipation regimes. The adiabatic elimination of the optical variable, valid in the normal dissipation regime $(\kappa\gg\gamma)$, makes the mechanical oscillator experience two-phonon absorption and emission processes with the optical reservoir ~\cite{PhysRevA.95.053844}. On the other hand, for the case of adiabatic elimination of the mechanical variable, valid in the reversed dissipation regime $(\gamma\gg\kappa)$, the cavity experiences one photon emission and absorption processes with the mechanical reservoir.

\section{Results}
\label{sec:Results}
It is well known that master equations with the form of Eq.~(\ref{eq:key_result}) are exactly solvable~\cite{Pierre_book}. Therefore all solutions at arbitrary times can be obtained analytically. For example, the equation of motion for the mean amplitude of the cavity field at time $t$ is given by
\begin{equation}
 \frac{d}{dt}\langle \hat a(t) \rangle= \left(i\Delta_c-\frac{\kappa_{\rm eff}}{2}\right)\langle \hat a(t) \rangle,
 \label{eq:field_amplitude}
\end{equation}
where we have, for typographical simplicity, defined the effective field decay rate as
\begin{equation}
 \kappa_{\rm eff} \equiv D_e-D_a = \kappa\left[1+C_2(2\bar n_m+1)\right]. 
 \label{eq:effective_decay_rate}
\end{equation}
Here, $C_2=\frac{8g_2^2}{\gamma\kappa}$ is the quadratic optomechanical cooperativity which is a measure of how strong the optomechanical system is coupled with respect to the two decay channels, i.e., optical and mechanical. Eq.~(\ref{eq:effective_decay_rate}) describes a broadening of the cavity linewidth due to the thermal and quantum fluctuations of the mechanical oscillator. Even when the mechanical bath is in zero temperature, the quantum fluctuations of the mechanics broadens the cavity linewidth by an amount of $\kappa C_2=\frac{8g_2^2}{\gamma}$. The added contribution of the thermal noise of the mechanics on the cavity linewidth broadening is proportional to the mean phonon number of the mechanics. 

\begin{figure}[]
\includegraphics[width=0.48\textwidth]{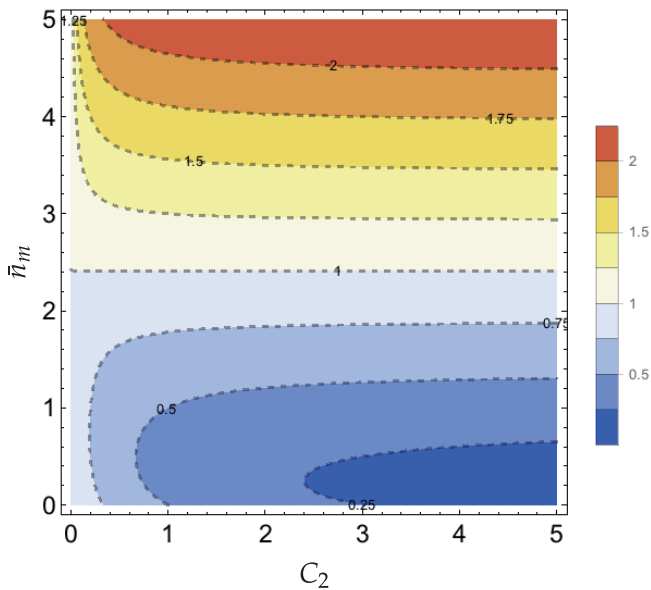}
\caption{Steady-state mean photon number of the cavity field as functions of the mean phonon number of the mechanical oscillator and the quadratic optomechanical cooperativity when $\bar n_o=1.0$. }
\label{fig:Mean_photon_number_quadratic}
\end{figure}

The equation of motion for the mean photon number is also given by 
\begin{equation}
 \frac{d}{dt}\langle \hat a^\dag(t) \hat a(t)\rangle=-\kappa_{\rm eff}\langle \hat a^\dag(t) \hat a(t)\rangle+D_a. 
 \label{eq:mean_photon_quadratic}
\end{equation}
Setting the left-hand-side of Eq.~(\ref{eq:mean_photon_quadratic}) to zero, one can find the steady-state mean photon number of the cavity field as 
\begin{equation}
\bar n_{\rm ss} = \frac{D_a}{\kappa_{\rm eff}}= \frac{\bar n_o }{1+C_2(2\bar n_m+1)}+\frac{ C_2\bar n_m^2}{1+C_2(2\bar n_m+1)} . 
\label{eq:ss_mean_photon_number_quadratic}
\end{equation}
Notice that the first term of Eq.~(\ref{eq:ss_mean_photon_number_quadratic}) decreases monotonically while the second term increases monotonically with increasing $\bar n_m$. The steady-state mean photon number becomes minimum when the mean phonon number of the mechanical oscillator is
\begin{equation}
 \bar n_m^{s} = \frac{-1-C_2+\sqrt{(1+C_2)^2+4C_2 \bar n_o}}{2C_2}. 
\end{equation}
Figure~\ref{fig:Mean_photon_number_quadratic} shows the steady-state mean photon number of the cavity field as functions of the mean phonon number of the mechanical heat bath as well as the quadratic optomechanical cooperativity. Interestingly, if the mean phonon number of the mechanical heat bath is equal to the critical value defined as
\begin{equation}
 \bar n_{m}^{c} = \bar n_o+\sqrt{\bar n_o(1+\bar n_o)},
 \label{eq:critical_number}
\end{equation}
the steady-state mean photon number of the cavity field is equal to that of the optical bath, $\bar n_{\rm ss}=\bar n_o$, independent of the optomechanical cooperativity. Furthermore, the cavity field in the steady state is being heated when $\bar n_m>\bar n_{m}^{c}$ and cooled when $\bar n_m<\bar n_{m}^{c}$. It should also be noted that the unity inside the square root in the right-hand-side of Eq.~(\ref{eq:critical_number}) is due to the quantum nature of the mechanical oscillator. 

\begin{figure}[]
\includegraphics[width=0.48\textwidth]{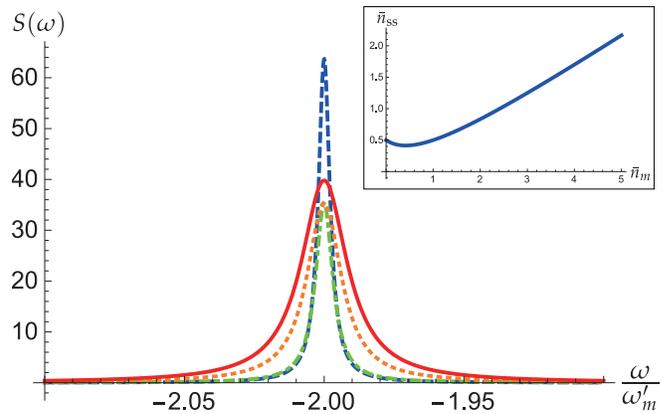}

\caption{Noise spectra of the cavity field for several representative mean phonon numbers of the mechanics: ${\bar n}_m=0$ (blue dashed curve), ${\bar n}_m=0.5$ (green dot-dashed curve), ${\bar n}_m=2.414$ (orange dotted curve), and ${\bar n}_m=3.6$ (red solid curve). Other parameters are $\kappa/\omega_m'=0.0025,~\bar n_o=1.0,~C_2=1.0$. The inset shows a plot of ${\bar n}_{\rm ss}$ as a function of ${\bar n}_m$ for the parameters $\bar n_o=1.0,~C_2=1.0$, describing the temperature of the cavity field with respect to that of the mechanical bath.}
\label{fig:Noise_spectrum_quadratic}
\end{figure}

Beyond the benefits of cooling a quantum system via reservoir engineering, a possible application of this system is for evaluating the temperature of the mechanical oscillator through linewidth measurement of the cavity field noise spectrum. The noise spectrum of the cavity field can be obtained by taking the Fourier transform of the two-time correlation function of the cavity field $ \langle \hat a^\dag(t+\tau)\hat a(t)\rangle$ in the long-time limit for $t$~\cite{Quantum_noise_book}. In the limit where the quantum noise is wide-sense-stationary, the noise spectrum of the cavity field is given by~\cite{Scully_book}
\begin{equation}
 S(\omega) = \frac{1}{\pi}{\rm Re}\int_{0}^{\infty}d\tau~ e^{i\omega \tau}\langle \hat a^\dag(t+\tau)\hat a(t)\rangle_{\rm ss} ,
 \label{eq:formal_power_spectrum}
\end{equation}
where $\rm ss$ indicates the fact that the cavity field is in steady-state.
Now that Eq.~(\ref{eq:field_amplitude}) is a linear differential equation for the field amplitude, one can make use of the quantum regression formula~\cite{Carmichael_book_1} to write
\begin{equation}
\frac{d}{d\tau} \langle \hat a^\dag(t+\tau) \hat a(t) \rangle =\left(-i\Delta_c-\frac{\kappa_{\rm eff}}{2}\right)\langle \hat a^\dag(t+\tau) \hat a(t) \rangle.
\label{eq:quantum_regression}
\end{equation}
Substituting the formal solution of Eq.~(\ref{eq:quantum_regression}) in the long time limit for $t$ into Eq.~(\ref{eq:formal_power_spectrum}) gives the noise spectrum of the cavity field
\begin{equation}
 S(\omega) =  \frac{\bar n_{\rm ss}}{\pi}\frac{\kappa_{\rm eff}/2}{(\Delta_c-\omega)^2+\kappa_{\rm eff}^2/4}.
 \label{eq:noise_spectrum}
\end{equation}
Notice that the field correlation function decays exponentially with $\tau$ and that the noise spectrum is a Lorentzian distribution centered at $\omega = \Delta_c=-2\omega_m'$ with full-width $\kappa_{\rm eff}$.
Figure~\ref{fig:Noise_spectrum_quadratic} shows the noise spectrum of the cavity field for a sample of representative mean phonon numbers of the mechanical oscillator. For simplicity, we assume that $\bar n_o=1.0$, $C_2=1.0$ which result in $\bar n_m^s = 0.414$, $\bar n_m^c = 2.414$. 
    
The broadening of the noise spectrum of the cavity field indicates the increase of the effective decay rate of the cavity field according to Eq.~(\ref{eq:effective_decay_rate}) : $\kappa_{\rm eff}/\omega_m' = 0.005$ (blue dashed curve), $\kappa_{\rm eff}/\omega_m' = 0.0075$ (green dotted curve), $\kappa_{\rm eff}/\omega_m' = 0.017$ (orange dot-dashed curve), $\kappa_{\rm eff}/\omega_m' = 0.023$ (red solid curve). It should be noted that the cavity linewidth is still broadened for the limiting case of a zero temperature mechanical reservoir, due to the quantum noise of the mechanics (blue dashed curve). To this end, the measurement of the noise spectrum, specifically, the broadening of the noise spectrum of the cavity field, allows one to estimate the temperature of the mechanical oscillator. Keeping in mind that the area under the noise spectrum of the cavity field is the same as the mean photon number of the cavity field~\cite{Pierre_book}, it is evident that the cavity field is cooled to a lower temperature when $\bar n_m=0.5$ (green dot-dashed curve) than when $\bar n_m=0$ (blue dashed curve). This behavior is depicted in the inset of Figure 3 where we plot the steady-state mean cavity photon number with respect to the thermal occupation number of the mechanical bath.
    
Finally, we remark that the cavity field described by the master equation of the form of Eq.~(\ref{eq:key_result}) approaches a thermal mixture with the second-order correlation function $g^{(2)}(0)\equiv\langle \hat a ^{\dag 2}(t)\hat a^2(t)\rangle_{\rm ss}/{\bar n}_{\rm ss}^2=2$.

\subsection{Comparison with the linear optomechanical system in the reversed dissipation regime}
In this subsection, we compare the behavior of the quadratic optomechanical system discussed above with that of a linear optomechanical system where the position of a mechanical oscillator linearly modifies the frequency of the cavity field in the reversed dissipation regime~\cite{PhysRevLett.113.023604, Toth:2017aa}. 

\begin{figure}[]
\includegraphics[width=0.48\textwidth]{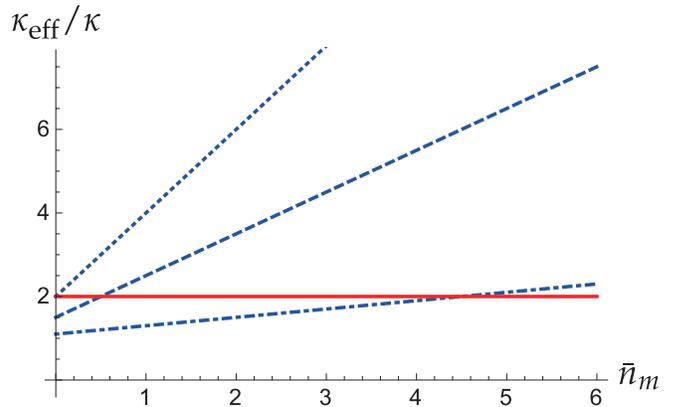}
\caption{Effective decay rate of the cavity field coupled to the mechanics linearly (red solid line) and quadratically (blue lines) as a function of the mean phonon number of the mechanics. 
For the linear coupling, we have used $C_1=1.0$ and for the quadratic coupling, we have used $C_2=0.1$ (blue dot-dashed line), $C_2=0.5$ (blue dashed line), $C_2=1.0$ (blue dotted line).
}
\label{fig:keff}
\end{figure}

One can exploit the same procedure for obtaining the effective master equation derived in the sections above. As expected, the master equation for the reduced density operator for the cavity field is of the form Eq.~(\ref{eq:key_result}) with corresponding emission $ D_e^{(L)}$ and absorption $ D_a^{(L)}$ coefficients1
\begin{eqnarray}
 D_e^{(L)} &=& \kappa({\bar n_o}+1) + \frac{4g_1^2}{\gamma}({\bar n_m}+1), \\
 D_a^{(L)}&=& \kappa{\bar n_o} + \frac{4g_1^2}{\gamma}{\bar n_m},
\end{eqnarray}
 where $g_1$ is the field-amplified optomechanical strength when the optomechanical system is operated in the resolved-sideband limit and the external pump is red-detuned by the mechanical frequency from the cavity resonance. In this case, the cavity linewidth can be expressed by
\begin{equation}
 \kappa_{\rm eff}^{(L)} = \kappa\left(1+C_1\right), 
\end{equation}
where $C_1=\frac{4g_1^2}{\gamma\kappa}$ is the linear optomechanical cooperativity. In contrast to the case of quadratic coupling, the effective decay rate of the cavity field for linear coupling does not involve the mean phonon number of the mechanical oscillator. This fact indicates that the thermal noise of the mechanics does not play a role in broadening the cavity linewidth. Rather, the cavity linewidth is broadened by increasing the optomechanical coupling strength $g_1$ and the mechanical damping rate $\gamma$. We plot the cavity field decay rate as a function of the mean phonon number of the mechanics for both linear and quadratic optomechanical systems in Figure~\ref{fig:keff}. Most notably, the cavity field linewidth for the linear optomechanics (red line) is constant with respect to $\bar n_{m}$, while that for quadratic optomechanics (blue lines) increases proportionally to $\bar n_{m}$ with a rate of increase determined by the cooperativity $C_{2}$.

\begin{figure}[]
\includegraphics[width=0.48\textwidth]{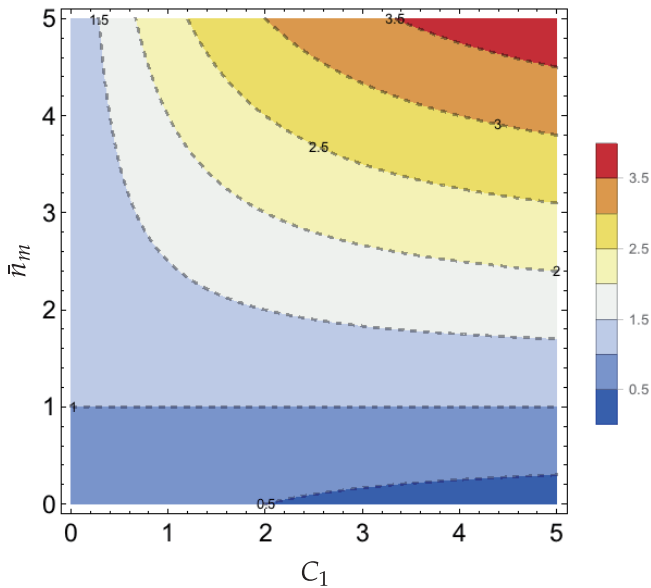}
\caption{Steady-state mean photon number of the cavity field coupled to the mechanics via the linear coupling. It is functions of the mean phonon number of the mechanical oscillator and the linear optomechanical cooperativity. The parameter we have used is $\bar n_o=1.0$.}
\label{fig:Mean_photon_number_linear}
\end{figure}

In the same manner, one can also find the mean photon number in the steady-state as
\begin{equation}
  \bar n_{\rm ss}^{(L)} = \frac{D_a^{(L)}}{\kappa_{\rm eff}^{(L)}}=\frac{\bar n_o }{1+C_1}+\frac{C_1\bar n_m}{1+C_1},
  \label{eq:linear_mean_photon}
\end{equation}
Notice that the first term of Eq.~(\ref{eq:linear_mean_photon}) is constant in the mean phonon number of the mechanics while the second term increases monotonically with increasing $\bar n_m$. It follows that the steady-state mean photon number becomes minimum when the mechanical oscillator is in contact with a reservoir at zero temperature, $\bar n_m^s=0$. 
For linear optomechanics, as expected, the steady-state mean photon number of the cavity field is equal to the mean photon number of the optical bath, irregardless of the optomechanical cooperativity, namely $\bar n_{\rm ss}^{(L)}=\bar n_o$, when the mean phonon number of the mechanical heat bath is
\begin{equation}
 \bar n_{m}^{c} = \bar n_o.
 \label{eq:critical_number2}
\end{equation}
 As usual, the cavity field in the steady state is always cooled (heated) when  $\bar n_m<\bar n_{m}^{c}$ ($\bar n_m>\bar n_{m}^{c}$). This behavior is illustrated in Figure~\ref{fig:Mean_photon_number_linear} which should be compared to the quadratically coupled case shown in Figure 2.
Note that in the case where the mechanics is in a reservoir at zero temperature in the reversed dissipation regime, the absorption and emission coefficients have the same dependence on their respective cooperativity for linear and quadratic coupling. 

Finally, the noise spectrum of the cavity field in this case is also Lorentzian of the form of Eq.~(\ref{eq:noise_spectrum}) with replacements $\kappa_{\rm eff}$ with $\kappa_{\rm eff}^{(L)}$ and $\bar n_{\rm ss}$ with $\bar n_{\rm ss}^{(L)}$. As expected, there is no broadening in the noise spectra with respect to the thermal occupation of the mechanics, but the area under the spectrum, i.e., the mean photon number increases with $\bar n_m$. 

\section{Conclusions}
\label{Sec:Conclusions}
We have theoretically examined the cavity electromagnetic field of an optomechanical resonator coupled to a mechanical reservoir with the mechanical damping rate much larger than the cavity field dissipation rate. In this reversed dissipation regime, we obtain analytic solutions to the steady-state mean photon number of the cavity field along with a description for the noise spectra. The parameter conditions for cooling the cavity field with the mechanical oscillator is presented particularly when the optomechanical interaction is nonlinear, quadratic. The results show that cooling of the cavity field can be enhanced with quadratic interaction at play, compared to the typical linear optomechanical interaction, within some regions in parameter space. By observing the cavity noise spectra for both linear and quadratic optomechanical interactions we have shown that the later display linewidths depending on the mean phonon number of the mechanics, demonstrating potential applications for precision thermometry~\cite{PhysRevA.93.013836}.  Notably, the theory explored in this paper can be applied to cavity fields in both microwave and optical regimes. Optomechanical devices in the reversed dissipation regime with optical cavity photons have not yet been demonstrated to our knowledge. However, optomechanical cooling has been experimentally shown with optical fields~\cite{Chan:2011aa}, and this capability gives rise to opening additional dissipation channels for the mechanical oscillator. By implementing a secondary optical cavity field with a small energy decay rate~\cite{PhysRevLett.113.023604}, the resultant optomechanical system can reside in the reversed dissipation regime providing possibilities to exhibit behavior presented in this paper.

\acknowledgments
This research was supported by the National Research Foundation of Korea (NRF) grant funded by the Korea government (MSIP) (No.~2015R1C1A1A01052349) and the R\&D Convergence Program of NST (National Research Council of Science and Technology) of the Republic of Korea (Grant No. CAP-15-08-KRISS). 

\appendix

\bibliography{Reversed_PRA}

\begin{thebibliography}{35}%
\makeatletter
\providecommand \@ifxundefined [1]{%
 \@ifx{#1\undefined}
}%
\providecommand \@ifnum [1]{%
 \ifnum #1\expandafter \@firstoftwo
 \else \expandafter \@secondoftwo
 \fi
}%
\providecommand \@ifx [1]{%
 \ifx #1\expandafter \@firstoftwo
 \else \expandafter \@secondoftwo
 \fi
}%
\providecommand \natexlab [1]{#1}%
\providecommand \enquote  [1]{``#1''}%
\providecommand \bibnamefont  [1]{#1}%
\providecommand \bibfnamefont [1]{#1}%
\providecommand \citenamefont [1]{#1}%
\providecommand \href@noop [0]{\@secondoftwo}%
\providecommand \href [0]{\begingroup \@sanitize@url \@href}%
\providecommand \@href[1]{\@@startlink{#1}\@@href}%
\providecommand \@@href[1]{\endgroup#1\@@endlink}%
\providecommand \@sanitize@url [0]{\catcode `\\12\catcode `\$12\catcode
  `\&12\catcode `\#12\catcode `\^12\catcode `\_12\catcode `\%12\relax}%
\providecommand \@@startlink[1]{}%
\providecommand \@@endlink[0]{}%
\providecommand \url  [0]{\begingroup\@sanitize@url \@url }%
\providecommand \@url [1]{\endgroup\@href {#1}{\urlprefix }}%
\providecommand \urlprefix  [0]{URL }%
\providecommand \Eprint [0]{\href }%
\providecommand \doibase [0]{http://dx.doi.org/}%
\providecommand \selectlanguage [0]{\@gobble}%
\providecommand \bibinfo  [0]{\@secondoftwo}%
\providecommand \bibfield  [0]{\@secondoftwo}%
\providecommand \translation [1]{[#1]}%
\providecommand \BibitemOpen [0]{}%
\providecommand \bibitemStop [0]{}%
\providecommand \bibitemNoStop [0]{.\EOS\space}%
\providecommand \EOS [0]{\spacefactor3000\relax}%
\providecommand \BibitemShut  [1]{\csname bibitem#1\endcsname}%
\let\auto@bib@innerbib\@empty
\bibitem [{\citenamefont {Meystre}(2013)}]{ANDP:ANDP201200226}%
  \BibitemOpen
  \bibfield  {author} {\bibinfo {author} {\bibfnamefont {P.}~\bibnamefont
  {Meystre}},\ }\href {\doibase 10.1002/andp.201200226} {\bibfield  {journal}
  {\bibinfo  {journal} {Annalen der Physik}\ }\textbf {\bibinfo {volume}
  {525}},\ \bibinfo {pages} {215} (\bibinfo {year} {2013})}\BibitemShut
  {NoStop}%
\bibitem [{\citenamefont {Aspelmeyer}\ \emph {et~al.}(2014)\citenamefont
  {Aspelmeyer}, \citenamefont {Kippenberg},\ and\ \citenamefont
  {Marquardt}}]{RevModPhys.86.1391}%
  \BibitemOpen
  \bibfield  {author} {\bibinfo {author} {\bibfnamefont {M.}~\bibnamefont
  {Aspelmeyer}}, \bibinfo {author} {\bibfnamefont {T.~J.}\ \bibnamefont
  {Kippenberg}}, \ and\ \bibinfo {author} {\bibfnamefont {F.}~\bibnamefont
  {Marquardt}},\ }\href {\doibase 10.1103/RevModPhys.86.1391} {\bibfield
  {journal} {\bibinfo  {journal} {Rev. Mod. Phys.}\ }\textbf {\bibinfo {volume}
  {86}},\ \bibinfo {pages} {1391} (\bibinfo {year} {2014})}\BibitemShut
  {NoStop}%
\bibitem [{\citenamefont {Bowen}\ and\ \citenamefont
  {Milburn}(2016)}]{Milburn_optomechanics_book}%
  \BibitemOpen
  \bibfield  {author} {\bibinfo {author} {\bibfnamefont {W.~P.}\ \bibnamefont
  {Bowen}}\ and\ \bibinfo {author} {\bibfnamefont {G.~J.}\ \bibnamefont
  {Milburn}},\ }\href@noop {} {\emph {\bibinfo {title} {Quantum
  Optomechanics}}}\ (\bibinfo  {publisher} {CRC Press, Boca Raton, FL},\
  \bibinfo {year} {2016})\BibitemShut {NoStop}%
\bibitem [{\citenamefont {Purdy}\ \emph
  {et~al.}(2013{\natexlab{a}})\citenamefont {Purdy}, \citenamefont {Peterson},\
  and\ \citenamefont {Regal}}]{Purdy801}%
  \BibitemOpen
  \bibfield  {author} {\bibinfo {author} {\bibfnamefont {T.~P.}\ \bibnamefont
  {Purdy}}, \bibinfo {author} {\bibfnamefont {R.~W.}\ \bibnamefont {Peterson}},
  \ and\ \bibinfo {author} {\bibfnamefont {C.~A.}\ \bibnamefont {Regal}},\
  }\href {\doibase 10.1126/science.1231282} {\bibfield  {journal} {\bibinfo
  {journal} {Science}\ }\textbf {\bibinfo {volume} {339}},\ \bibinfo {pages}
  {801} (\bibinfo {year} {2013}{\natexlab{a}})}\BibitemShut {NoStop}%
\bibitem [{\citenamefont {Schreppler}\ \emph {et~al.}(2014)\citenamefont
  {Schreppler}, \citenamefont {Spethmann}, \citenamefont {Brahms},
  \citenamefont {Botter}, \citenamefont {Barrios},\ and\ \citenamefont
  {Stamper-Kurn}}]{Schreppler1486}%
  \BibitemOpen
  \bibfield  {author} {\bibinfo {author} {\bibfnamefont {S.}~\bibnamefont
  {Schreppler}}, \bibinfo {author} {\bibfnamefont {N.}~\bibnamefont
  {Spethmann}}, \bibinfo {author} {\bibfnamefont {N.}~\bibnamefont {Brahms}},
  \bibinfo {author} {\bibfnamefont {T.}~\bibnamefont {Botter}}, \bibinfo
  {author} {\bibfnamefont {M.}~\bibnamefont {Barrios}}, \ and\ \bibinfo
  {author} {\bibfnamefont {D.~M.}\ \bibnamefont {Stamper-Kurn}},\ }\href
  {\doibase 10.1126/science.1249850} {\bibfield  {journal} {\bibinfo  {journal}
  {Science}\ }\textbf {\bibinfo {volume} {344}},\ \bibinfo {pages} {1486}
  (\bibinfo {year} {2014})}\BibitemShut {NoStop}%
\bibitem [{\citenamefont {Romero-Isart}(2011)}]{PhysRevA.84.052121}%
  \BibitemOpen
  \bibfield  {author} {\bibinfo {author} {\bibfnamefont {O.}~\bibnamefont
  {Romero-Isart}},\ }\href {\doibase 10.1103/PhysRevA.84.052121} {\bibfield
  {journal} {\bibinfo  {journal} {Phys. Rev. A}\ }\textbf {\bibinfo {volume}
  {84}},\ \bibinfo {pages} {052121} (\bibinfo {year} {2011})}\BibitemShut
  {NoStop}%
\bibitem [{\citenamefont {Camerer}\ \emph {et~al.}(2011)\citenamefont
  {Camerer}, \citenamefont {Korppi}, \citenamefont {J\"ockel}, \citenamefont
  {Hunger}, \citenamefont {H\"ansch},\ and\ \citenamefont
  {Treutlein}}]{PhysRevLett.107.223001}%
  \BibitemOpen
  \bibfield  {author} {\bibinfo {author} {\bibfnamefont {S.}~\bibnamefont
  {Camerer}}, \bibinfo {author} {\bibfnamefont {M.}~\bibnamefont {Korppi}},
  \bibinfo {author} {\bibfnamefont {A.}~\bibnamefont {J\"ockel}}, \bibinfo
  {author} {\bibfnamefont {D.}~\bibnamefont {Hunger}}, \bibinfo {author}
  {\bibfnamefont {T.~W.}\ \bibnamefont {H\"ansch}}, \ and\ \bibinfo {author}
  {\bibfnamefont {P.}~\bibnamefont {Treutlein}},\ }\href {\doibase
  10.1103/PhysRevLett.107.223001} {\bibfield  {journal} {\bibinfo  {journal}
  {Phys. Rev. Lett.}\ }\textbf {\bibinfo {volume} {107}},\ \bibinfo {pages}
  {223001} (\bibinfo {year} {2011})}\BibitemShut {NoStop}%
\bibitem [{\citenamefont {Dar\'azs}\ \emph {et~al.}(2014)\citenamefont
  {Dar\'azs}, \citenamefont {Kurucz}, \citenamefont {K\'alm\'an}, \citenamefont
  {Kiss}, \citenamefont {Fort\'agh},\ and\ \citenamefont
  {Domokos}}]{PhysRevLett.112.133603}%
  \BibitemOpen
  \bibfield  {author} {\bibinfo {author} {\bibfnamefont {Z.}~\bibnamefont
  {Dar\'azs}}, \bibinfo {author} {\bibfnamefont {Z.}~\bibnamefont {Kurucz}},
  \bibinfo {author} {\bibfnamefont {O.}~\bibnamefont {K\'alm\'an}}, \bibinfo
  {author} {\bibfnamefont {T.}~\bibnamefont {Kiss}}, \bibinfo {author}
  {\bibfnamefont {J.}~\bibnamefont {Fort\'agh}}, \ and\ \bibinfo {author}
  {\bibfnamefont {P.}~\bibnamefont {Domokos}},\ }\href {\doibase
  10.1103/PhysRevLett.112.133603} {\bibfield  {journal} {\bibinfo  {journal}
  {Phys. Rev. Lett.}\ }\textbf {\bibinfo {volume} {112}},\ \bibinfo {pages}
  {133603} (\bibinfo {year} {2014})}\BibitemShut {NoStop}%
\bibitem [{\citenamefont {Teufel}\ \emph {et~al.}(2011)\citenamefont {Teufel},
  \citenamefont {Donner}, \citenamefont {Li}, \citenamefont {Harlow},
  \citenamefont {Allman}, \citenamefont {Cicak}, \citenamefont {Sirois},
  \citenamefont {Whittaker}, \citenamefont {Lehnert},\ and\ \citenamefont
  {Simmonds}}]{Teufel:2011aa}%
  \BibitemOpen
  \bibfield  {author} {\bibinfo {author} {\bibfnamefont {J.~D.}\ \bibnamefont
  {Teufel}}, \bibinfo {author} {\bibfnamefont {T.}~\bibnamefont {Donner}},
  \bibinfo {author} {\bibfnamefont {D.}~\bibnamefont {Li}}, \bibinfo {author}
  {\bibfnamefont {J.~W.}\ \bibnamefont {Harlow}}, \bibinfo {author}
  {\bibfnamefont {M.~S.}\ \bibnamefont {Allman}}, \bibinfo {author}
  {\bibfnamefont {K.}~\bibnamefont {Cicak}}, \bibinfo {author} {\bibfnamefont
  {A.~J.}\ \bibnamefont {Sirois}}, \bibinfo {author} {\bibfnamefont {J.~D.}\
  \bibnamefont {Whittaker}}, \bibinfo {author} {\bibfnamefont {K.~W.}\
  \bibnamefont {Lehnert}}, \ and\ \bibinfo {author} {\bibfnamefont {R.~W.}\
  \bibnamefont {Simmonds}},\ }\href {http://dx.doi.org/10.1038/nature10261}
  {\bibfield  {journal} {\bibinfo  {journal} {Nature}\ }\textbf {\bibinfo
  {volume} {475}},\ \bibinfo {pages} {359} (\bibinfo {year}
  {2011})}\BibitemShut {NoStop}%
\bibitem [{\citenamefont {Chan}\ \emph {et~al.}(2011)\citenamefont {Chan},
  \citenamefont {Alegre}, \citenamefont {Safavi-Naeini}, \citenamefont {Hill},
  \citenamefont {Krause}, \citenamefont {Groblacher}, \citenamefont
  {Aspelmeyer},\ and\ \citenamefont {Painter}}]{Chan:2011aa}%
  \BibitemOpen
  \bibfield  {author} {\bibinfo {author} {\bibfnamefont {J.}~\bibnamefont
  {Chan}}, \bibinfo {author} {\bibfnamefont {T.~P.~M.}\ \bibnamefont {Alegre}},
  \bibinfo {author} {\bibfnamefont {A.~H.}\ \bibnamefont {Safavi-Naeini}},
  \bibinfo {author} {\bibfnamefont {J.~T.}\ \bibnamefont {Hill}}, \bibinfo
  {author} {\bibfnamefont {A.}~\bibnamefont {Krause}}, \bibinfo {author}
  {\bibfnamefont {S.}~\bibnamefont {Groblacher}}, \bibinfo {author}
  {\bibfnamefont {M.}~\bibnamefont {Aspelmeyer}}, \ and\ \bibinfo {author}
  {\bibfnamefont {O.}~\bibnamefont {Painter}},\ }\href
  {http://dx.doi.org/10.1038/nature10461} {\bibfield  {journal} {\bibinfo
  {journal} {Nature}\ }\textbf {\bibinfo {volume} {478}},\ \bibinfo {pages}
  {89} (\bibinfo {year} {2011})}\BibitemShut {NoStop}%
\bibitem [{\citenamefont {Agarwal}\ and\ \citenamefont
  {Huang}(2010)}]{PhysRevA.81.041803}%
  \BibitemOpen
  \bibfield  {author} {\bibinfo {author} {\bibfnamefont {G.~S.}\ \bibnamefont
  {Agarwal}}\ and\ \bibinfo {author} {\bibfnamefont {S.}~\bibnamefont
  {Huang}},\ }\href {\doibase 10.1103/PhysRevA.81.041803} {\bibfield  {journal}
  {\bibinfo  {journal} {Phys. Rev. A}\ }\textbf {\bibinfo {volume} {81}},\
  \bibinfo {pages} {041803} (\bibinfo {year} {2010})}\BibitemShut {NoStop}%
\bibitem [{\citenamefont {Weis}\ \emph {et~al.}(2010)\citenamefont {Weis},
  \citenamefont {Rivi{\`e}re}, \citenamefont {Del{\'e}glise}, \citenamefont
  {Gavartin}, \citenamefont {Arcizet}, \citenamefont {Schliesser},\ and\
  \citenamefont {Kippenberg}}]{Weis1520}%
  \BibitemOpen
  \bibfield  {author} {\bibinfo {author} {\bibfnamefont {S.}~\bibnamefont
  {Weis}}, \bibinfo {author} {\bibfnamefont {R.}~\bibnamefont {Rivi{\`e}re}},
  \bibinfo {author} {\bibfnamefont {S.}~\bibnamefont {Del{\'e}glise}}, \bibinfo
  {author} {\bibfnamefont {E.}~\bibnamefont {Gavartin}}, \bibinfo {author}
  {\bibfnamefont {O.}~\bibnamefont {Arcizet}}, \bibinfo {author} {\bibfnamefont
  {A.}~\bibnamefont {Schliesser}}, \ and\ \bibinfo {author} {\bibfnamefont
  {T.~J.}\ \bibnamefont {Kippenberg}},\ }\href {\doibase
  10.1126/science.1195596} {\bibfield  {journal} {\bibinfo  {journal}
  {Science}\ }\textbf {\bibinfo {volume} {330}},\ \bibinfo {pages} {1520}
  (\bibinfo {year} {2010})}\BibitemShut {NoStop}%
\bibitem [{\citenamefont {Verhagen}\ \emph {et~al.}(2012)\citenamefont
  {Verhagen}, \citenamefont {Deleglise}, \citenamefont {Weis}, \citenamefont
  {Schliesser},\ and\ \citenamefont {Kippenberg}}]{Verhagen:2012aa}%
  \BibitemOpen
  \bibfield  {author} {\bibinfo {author} {\bibfnamefont {E.}~\bibnamefont
  {Verhagen}}, \bibinfo {author} {\bibfnamefont {S.}~\bibnamefont {Deleglise}},
  \bibinfo {author} {\bibfnamefont {S.}~\bibnamefont {Weis}}, \bibinfo {author}
  {\bibfnamefont {A.}~\bibnamefont {Schliesser}}, \ and\ \bibinfo {author}
  {\bibfnamefont {T.~J.}\ \bibnamefont {Kippenberg}},\ }\href
  {http://dx.doi.org/10.1038/nature10787} {\bibfield  {journal} {\bibinfo
  {journal} {Nature}\ }\textbf {\bibinfo {volume} {482}},\ \bibinfo {pages}
  {63} (\bibinfo {year} {2012})}\BibitemShut {NoStop}%
\bibitem [{\citenamefont {Palomaki}\ \emph {et~al.}(2013)\citenamefont
  {Palomaki}, \citenamefont {Teufel}, \citenamefont {Simmonds},\ and\
  \citenamefont {Lehnert}}]{Palomaki710}%
  \BibitemOpen
  \bibfield  {author} {\bibinfo {author} {\bibfnamefont {T.~A.}\ \bibnamefont
  {Palomaki}}, \bibinfo {author} {\bibfnamefont {J.~D.}\ \bibnamefont
  {Teufel}}, \bibinfo {author} {\bibfnamefont {R.~W.}\ \bibnamefont
  {Simmonds}}, \ and\ \bibinfo {author} {\bibfnamefont {K.~W.}\ \bibnamefont
  {Lehnert}},\ }\href {\doibase 10.1126/science.1244563} {\bibfield  {journal}
  {\bibinfo  {journal} {Science}\ }\textbf {\bibinfo {volume} {342}},\ \bibinfo
  {pages} {710} (\bibinfo {year} {2013})}\BibitemShut {NoStop}%
\bibitem [{\citenamefont {Shkarin}\ \emph {et~al.}(2014)\citenamefont
  {Shkarin}, \citenamefont {Flowers-Jacobs}, \citenamefont {Hoch},
  \citenamefont {Kashkanova}, \citenamefont {Deutsch}, \citenamefont
  {Reichel},\ and\ \citenamefont {Harris}}]{PhysRevLett.112.013602}%
  \BibitemOpen
  \bibfield  {author} {\bibinfo {author} {\bibfnamefont {A.~B.}\ \bibnamefont
  {Shkarin}}, \bibinfo {author} {\bibfnamefont {N.~E.}\ \bibnamefont
  {Flowers-Jacobs}}, \bibinfo {author} {\bibfnamefont {S.~W.}\ \bibnamefont
  {Hoch}}, \bibinfo {author} {\bibfnamefont {A.~D.}\ \bibnamefont
  {Kashkanova}}, \bibinfo {author} {\bibfnamefont {C.}~\bibnamefont {Deutsch}},
  \bibinfo {author} {\bibfnamefont {J.}~\bibnamefont {Reichel}}, \ and\
  \bibinfo {author} {\bibfnamefont {J.~G.~E.}\ \bibnamefont {Harris}},\ }\href
  {\doibase 10.1103/PhysRevLett.112.013602} {\bibfield  {journal} {\bibinfo
  {journal} {Phys. Rev. Lett.}\ }\textbf {\bibinfo {volume} {112}},\ \bibinfo
  {pages} {013602} (\bibinfo {year} {2014})}\BibitemShut {NoStop}%
\bibitem [{\citenamefont {Spethmann}\ \emph {et~al.}(2016)\citenamefont
  {Spethmann}, \citenamefont {Kohler}, \citenamefont {Schreppler},
  \citenamefont {Buchmann},\ and\ \citenamefont
  {Stamper-Kurn}}]{Spethmann:2016aa}%
  \BibitemOpen
  \bibfield  {author} {\bibinfo {author} {\bibfnamefont {N.}~\bibnamefont
  {Spethmann}}, \bibinfo {author} {\bibfnamefont {J.}~\bibnamefont {Kohler}},
  \bibinfo {author} {\bibfnamefont {S.}~\bibnamefont {Schreppler}}, \bibinfo
  {author} {\bibfnamefont {L.}~\bibnamefont {Buchmann}}, \ and\ \bibinfo
  {author} {\bibfnamefont {D.~M.}\ \bibnamefont {Stamper-Kurn}},\ }\href
  {http://dx.doi.org/10.1038/nphys3515} {\bibfield  {journal} {\bibinfo
  {journal} {Nat Phys}\ }\textbf {\bibinfo {volume} {12}},\ \bibinfo {pages}
  {27} (\bibinfo {year} {2016})}\BibitemShut {NoStop}%
\bibitem [{\citenamefont {Brooks}\ \emph {et~al.}(2012)\citenamefont {Brooks},
  \citenamefont {Botter}, \citenamefont {Schreppler}, \citenamefont {Purdy},
  \citenamefont {Brahms},\ and\ \citenamefont {Stamper-Kurn}}]{Brooks:2012aa}%
  \BibitemOpen
  \bibfield  {author} {\bibinfo {author} {\bibfnamefont {D.~W.~C.}\
  \bibnamefont {Brooks}}, \bibinfo {author} {\bibfnamefont {T.}~\bibnamefont
  {Botter}}, \bibinfo {author} {\bibfnamefont {S.}~\bibnamefont {Schreppler}},
  \bibinfo {author} {\bibfnamefont {T.~P.}\ \bibnamefont {Purdy}}, \bibinfo
  {author} {\bibfnamefont {N.}~\bibnamefont {Brahms}}, \ and\ \bibinfo {author}
  {\bibfnamefont {D.~M.}\ \bibnamefont {Stamper-Kurn}},\ }\href
  {http://dx.doi.org/10.1038/nature11325} {\bibfield  {journal} {\bibinfo
  {journal} {Nature}\ }\textbf {\bibinfo {volume} {488}},\ \bibinfo {pages}
  {476} (\bibinfo {year} {2012})}\BibitemShut {NoStop}%
\bibitem [{\citenamefont {Safavi-Naeini}\ \emph {et~al.}(2013)\citenamefont
  {Safavi-Naeini}, \citenamefont {Groblacher}, \citenamefont {Hill},
  \citenamefont {Chan}, \citenamefont {Aspelmeyer},\ and\ \citenamefont
  {Painter}}]{Safavi-Naeini:2013aa}%
  \BibitemOpen
  \bibfield  {author} {\bibinfo {author} {\bibfnamefont {A.~H.}\ \bibnamefont
  {Safavi-Naeini}}, \bibinfo {author} {\bibfnamefont {S.}~\bibnamefont
  {Groblacher}}, \bibinfo {author} {\bibfnamefont {J.~T.}\ \bibnamefont
  {Hill}}, \bibinfo {author} {\bibfnamefont {J.}~\bibnamefont {Chan}}, \bibinfo
  {author} {\bibfnamefont {M.}~\bibnamefont {Aspelmeyer}}, \ and\ \bibinfo
  {author} {\bibfnamefont {O.}~\bibnamefont {Painter}},\ }\href
  {http://dx.doi.org/10.1038/nature12307} {\bibfield  {journal} {\bibinfo
  {journal} {Nature}\ }\textbf {\bibinfo {volume} {500}},\ \bibinfo {pages}
  {185} (\bibinfo {year} {2013})}\BibitemShut {NoStop}%
\bibitem [{\citenamefont {Purdy}\ \emph
  {et~al.}(2013{\natexlab{b}})\citenamefont {Purdy}, \citenamefont {Yu},
  \citenamefont {Peterson}, \citenamefont {Kampel},\ and\ \citenamefont
  {Regal}}]{PhysRevX.3.031012}%
  \BibitemOpen
  \bibfield  {author} {\bibinfo {author} {\bibfnamefont {T.~P.}\ \bibnamefont
  {Purdy}}, \bibinfo {author} {\bibfnamefont {P.-L.}\ \bibnamefont {Yu}},
  \bibinfo {author} {\bibfnamefont {R.~W.}\ \bibnamefont {Peterson}}, \bibinfo
  {author} {\bibfnamefont {N.~S.}\ \bibnamefont {Kampel}}, \ and\ \bibinfo
  {author} {\bibfnamefont {C.~A.}\ \bibnamefont {Regal}},\ }\href {\doibase
  10.1103/PhysRevX.3.031012} {\bibfield  {journal} {\bibinfo  {journal} {Phys.
  Rev. X}\ }\textbf {\bibinfo {volume} {3}},\ \bibinfo {pages} {031012}
  (\bibinfo {year} {2013}{\natexlab{b}})}\BibitemShut {NoStop}%
\bibitem [{\citenamefont {Poyatos}\ \emph {et~al.}(1996)\citenamefont
  {Poyatos}, \citenamefont {Cirac},\ and\ \citenamefont
  {Zoller}}]{PhysRevLett.77.4728}%
  \BibitemOpen
  \bibfield  {author} {\bibinfo {author} {\bibfnamefont {J.~F.}\ \bibnamefont
  {Poyatos}}, \bibinfo {author} {\bibfnamefont {J.~I.}\ \bibnamefont {Cirac}},
  \ and\ \bibinfo {author} {\bibfnamefont {P.}~\bibnamefont {Zoller}},\ }\href
  {\doibase 10.1103/PhysRevLett.77.4728} {\bibfield  {journal} {\bibinfo
  {journal} {Phys. Rev. Lett.}\ }\textbf {\bibinfo {volume} {77}},\ \bibinfo
  {pages} {4728} (\bibinfo {year} {1996})}\BibitemShut {NoStop}%
\bibitem [{\citenamefont {Krauter}\ \emph {et~al.}(2011)\citenamefont
  {Krauter}, \citenamefont {Muschik}, \citenamefont {Jensen}, \citenamefont
  {Wasilewski}, \citenamefont {Petersen}, \citenamefont {Cirac},\ and\
  \citenamefont {Polzik}}]{PhysRevLett.107.080503}%
  \BibitemOpen
  \bibfield  {author} {\bibinfo {author} {\bibfnamefont {H.}~\bibnamefont
  {Krauter}}, \bibinfo {author} {\bibfnamefont {C.~A.}\ \bibnamefont
  {Muschik}}, \bibinfo {author} {\bibfnamefont {K.}~\bibnamefont {Jensen}},
  \bibinfo {author} {\bibfnamefont {W.}~\bibnamefont {Wasilewski}}, \bibinfo
  {author} {\bibfnamefont {J.~M.}\ \bibnamefont {Petersen}}, \bibinfo {author}
  {\bibfnamefont {J.~I.}\ \bibnamefont {Cirac}}, \ and\ \bibinfo {author}
  {\bibfnamefont {E.~S.}\ \bibnamefont {Polzik}},\ }\href {\doibase
  10.1103/PhysRevLett.107.080503} {\bibfield  {journal} {\bibinfo  {journal}
  {Phys. Rev. Lett.}\ }\textbf {\bibinfo {volume} {107}},\ \bibinfo {pages}
  {080503} (\bibinfo {year} {2011})}\BibitemShut {NoStop}%
\bibitem [{\citenamefont {Kienzler}\ \emph {et~al.}(2015)\citenamefont
  {Kienzler}, \citenamefont {Lo}, \citenamefont {Keitch}, \citenamefont
  {de~Clercq}, \citenamefont {Leupold}, \citenamefont {Lindenfelser},
  \citenamefont {Marinelli}, \citenamefont {Negnevitsky},\ and\ \citenamefont
  {Home}}]{Kienzler53}%
  \BibitemOpen
  \bibfield  {author} {\bibinfo {author} {\bibfnamefont {D.}~\bibnamefont
  {Kienzler}}, \bibinfo {author} {\bibfnamefont {H.-Y.}\ \bibnamefont {Lo}},
  \bibinfo {author} {\bibfnamefont {B.}~\bibnamefont {Keitch}}, \bibinfo
  {author} {\bibfnamefont {L.}~\bibnamefont {de~Clercq}}, \bibinfo {author}
  {\bibfnamefont {F.}~\bibnamefont {Leupold}}, \bibinfo {author} {\bibfnamefont
  {F.}~\bibnamefont {Lindenfelser}}, \bibinfo {author} {\bibfnamefont
  {M.}~\bibnamefont {Marinelli}}, \bibinfo {author} {\bibfnamefont
  {V.}~\bibnamefont {Negnevitsky}}, \ and\ \bibinfo {author} {\bibfnamefont
  {J.~P.}\ \bibnamefont {Home}},\ }\href {\doibase 10.1126/science.1261033}
  {\bibfield  {journal} {\bibinfo  {journal} {Science}\ }\textbf {\bibinfo
  {volume} {347}},\ \bibinfo {pages} {53} (\bibinfo {year} {2015})}\BibitemShut
  {NoStop}%
\bibitem [{\citenamefont {Nunnenkamp}\ \emph {et~al.}(2014)\citenamefont
  {Nunnenkamp}, \citenamefont {Sudhir}, \citenamefont {Feofanov}, \citenamefont
  {Roulet},\ and\ \citenamefont {Kippenberg}}]{PhysRevLett.113.023604}%
  \BibitemOpen
  \bibfield  {author} {\bibinfo {author} {\bibfnamefont {A.}~\bibnamefont
  {Nunnenkamp}}, \bibinfo {author} {\bibfnamefont {V.}~\bibnamefont {Sudhir}},
  \bibinfo {author} {\bibfnamefont {A.~K.}\ \bibnamefont {Feofanov}}, \bibinfo
  {author} {\bibfnamefont {A.}~\bibnamefont {Roulet}}, \ and\ \bibinfo {author}
  {\bibfnamefont {T.~J.}\ \bibnamefont {Kippenberg}},\ }\href {\doibase
  10.1103/PhysRevLett.113.023604} {\bibfield  {journal} {\bibinfo  {journal}
  {Phys. Rev. Lett.}\ }\textbf {\bibinfo {volume} {113}},\ \bibinfo {pages}
  {023604} (\bibinfo {year} {2014})}\BibitemShut {NoStop}%
\bibitem [{\citenamefont {Toth}\ \emph {et~al.}(2017)\citenamefont {Toth},
  \citenamefont {Bernier}, \citenamefont {Nunnenkamp}, \citenamefont
  {Feofanov},\ and\ \citenamefont {Kippenberg}}]{Toth:2017aa}%
  \BibitemOpen
  \bibfield  {author} {\bibinfo {author} {\bibfnamefont {L.~D.}\ \bibnamefont
  {Toth}}, \bibinfo {author} {\bibfnamefont {N.~R.}\ \bibnamefont {Bernier}},
  \bibinfo {author} {\bibfnamefont {A.}~\bibnamefont {Nunnenkamp}}, \bibinfo
  {author} {\bibfnamefont {A.~K.}\ \bibnamefont {Feofanov}}, \ and\ \bibinfo
  {author} {\bibfnamefont {T.~J.}\ \bibnamefont {Kippenberg}},\ }\href
  {http://dx.doi.org/10.1038/nphys4121} {\bibfield  {journal} {\bibinfo
  {journal} {Nat Phys}\ }\textbf {\bibinfo {volume} {13}},\ \bibinfo {pages}
  {787} (\bibinfo {year} {2017})}\BibitemShut {NoStop}%
\bibitem [{\citenamefont {Nunnenkamp}\ \emph {et~al.}(2010)\citenamefont
  {Nunnenkamp}, \citenamefont {B\o{}rkje}, \citenamefont {Harris},\ and\
  \citenamefont {Girvin}}]{PhysRevA.82.021806}%
  \BibitemOpen
  \bibfield  {author} {\bibinfo {author} {\bibfnamefont {A.}~\bibnamefont
  {Nunnenkamp}}, \bibinfo {author} {\bibfnamefont {K.}~\bibnamefont
  {B\o{}rkje}}, \bibinfo {author} {\bibfnamefont {J.~G.~E.}\ \bibnamefont
  {Harris}}, \ and\ \bibinfo {author} {\bibfnamefont {S.~M.}\ \bibnamefont
  {Girvin}},\ }\href {\doibase 10.1103/PhysRevA.82.021806} {\bibfield
  {journal} {\bibinfo  {journal} {Phys. Rev. A}\ }\textbf {\bibinfo {volume}
  {82}},\ \bibinfo {pages} {021806} (\bibinfo {year} {2010})}\BibitemShut
  {NoStop}%
\bibitem [{\citenamefont {Sankey}\ \emph {et~al.}(2010)\citenamefont {Sankey},
  \citenamefont {Yang}, \citenamefont {Zwickl}, \citenamefont {Jayich},\ and\
  \citenamefont {Harris}}]{Sankey:2010aa}%
  \BibitemOpen
  \bibfield  {author} {\bibinfo {author} {\bibfnamefont {J.~C.}\ \bibnamefont
  {Sankey}}, \bibinfo {author} {\bibfnamefont {C.}~\bibnamefont {Yang}},
  \bibinfo {author} {\bibfnamefont {B.~M.}\ \bibnamefont {Zwickl}}, \bibinfo
  {author} {\bibfnamefont {A.~M.}\ \bibnamefont {Jayich}}, \ and\ \bibinfo
  {author} {\bibfnamefont {J.~G.~E.}\ \bibnamefont {Harris}},\ }\href
  {http://dx.doi.org/10.1038/nphys1707} {\bibfield  {journal} {\bibinfo
  {journal} {Nat Phys}\ }\textbf {\bibinfo {volume} {6}},\ \bibinfo {pages}
  {707} (\bibinfo {year} {2010})}\BibitemShut {NoStop}%
\bibitem [{\citenamefont {Para\"{\i}so}\ \emph {et~al.}(2015)\citenamefont
  {Para\"{\i}so}, \citenamefont {Kalaee}, \citenamefont {Zang}, \citenamefont
  {Pfeifer}, \citenamefont {Marquardt},\ and\ \citenamefont
  {Painter}}]{PhysRevX.5.041024}%
  \BibitemOpen
  \bibfield  {author} {\bibinfo {author} {\bibfnamefont {T.~K.}\ \bibnamefont
  {Para\"{\i}so}}, \bibinfo {author} {\bibfnamefont {M.}~\bibnamefont
  {Kalaee}}, \bibinfo {author} {\bibfnamefont {L.}~\bibnamefont {Zang}},
  \bibinfo {author} {\bibfnamefont {H.}~\bibnamefont {Pfeifer}}, \bibinfo
  {author} {\bibfnamefont {F.}~\bibnamefont {Marquardt}}, \ and\ \bibinfo
  {author} {\bibfnamefont {O.}~\bibnamefont {Painter}},\ }\href {\doibase
  10.1103/PhysRevX.5.041024} {\bibfield  {journal} {\bibinfo  {journal} {Phys.
  Rev. X}\ }\textbf {\bibinfo {volume} {5}},\ \bibinfo {pages} {041024}
  (\bibinfo {year} {2015})}\BibitemShut {NoStop}%
\bibitem [{\citenamefont {Meystre}\ and\ \citenamefont
  {Sargent~III}(2007)}]{Pierre_book}%
  \BibitemOpen
  \bibfield  {author} {\bibinfo {author} {\bibfnamefont {P.}~\bibnamefont
  {Meystre}}\ and\ \bibinfo {author} {\bibfnamefont {M.}~\bibnamefont
  {Sargent~III}},\ }\href@noop {} {\emph {\bibinfo {title} {Elements of Quantum
  Optics}}}\ (\bibinfo  {publisher} {Springer-Verlag Berlin Heidelberg},\
  \bibinfo {year} {2007})\BibitemShut {NoStop}%
\bibitem [{\citenamefont {Carmichael}(1999)}]{Carmichael_book_1}%
  \BibitemOpen
  \bibfield  {author} {\bibinfo {author} {\bibfnamefont {H.~J.}\ \bibnamefont
  {Carmichael}},\ }\href@noop {} {\emph {\bibinfo {title} {Statistical methods
  in quantum optics 1}}}\ (\bibinfo  {publisher} {Springer-Verlag, Berlin},\
  \bibinfo {year} {1999})\BibitemShut {NoStop}%
\bibitem [{\citenamefont {Walls}\ and\ \citenamefont
  {Milburn}(2008)}]{Milburn_book_1}%
  \BibitemOpen
  \bibfield  {author} {\bibinfo {author} {\bibfnamefont {D.~F.}\ \bibnamefont
  {Walls}}\ and\ \bibinfo {author} {\bibfnamefont {G.~J.}\ \bibnamefont
  {Milburn}},\ }\href@noop {} {\emph {\bibinfo {title} {Quantum Optics}}}\
  (\bibinfo  {publisher} {Springer-Verlag, Berlin},\ \bibinfo {year}
  {2008})\BibitemShut {NoStop}%
\bibitem [{\citenamefont {Carmichael}(2008)}]{Carmichael_book_2}%
  \BibitemOpen
  \bibfield  {author} {\bibinfo {author} {\bibfnamefont {H.~J.}\ \bibnamefont
  {Carmichael}},\ }\href@noop {} {\emph {\bibinfo {title} {Statistical methods
  in quantum optics 2}}}\ (\bibinfo  {publisher} {Springer-Verlag, Berlin},\
  \bibinfo {year} {2008})\BibitemShut {NoStop}%
\bibitem [{\citenamefont {Seok}\ and\ \citenamefont
  {Wright}(2017)}]{PhysRevA.95.053844}%
  \BibitemOpen
  \bibfield  {author} {\bibinfo {author} {\bibfnamefont {H.}~\bibnamefont
  {Seok}}\ and\ \bibinfo {author} {\bibfnamefont {E.~M.}\ \bibnamefont
  {Wright}},\ }\href {\doibase 10.1103/PhysRevA.95.053844} {\bibfield
  {journal} {\bibinfo  {journal} {Phys. Rev. A}\ }\textbf {\bibinfo {volume}
  {95}},\ \bibinfo {pages} {053844} (\bibinfo {year} {2017})}\BibitemShut
  {NoStop}%
\bibitem [{\citenamefont {Gardiner}\ and\ \citenamefont
  {Zoller}(2004)}]{Quantum_noise_book}%
  \BibitemOpen
  \bibfield  {author} {\bibinfo {author} {\bibfnamefont {C.}~\bibnamefont
  {Gardiner}}\ and\ \bibinfo {author} {\bibfnamefont {P.}~\bibnamefont
  {Zoller}},\ }\href@noop {} {\emph {\bibinfo {title} {Quantum Noise}}}\
  (\bibinfo  {publisher} {Springer-Verlag, Berlin},\ \bibinfo {year}
  {2004})\BibitemShut {NoStop}%
\bibitem [{\citenamefont {Scully}\ and\ \citenamefont
  {Zubairy}(2012)}]{Scully_book}%
  \BibitemOpen
  \bibfield  {author} {\bibinfo {author} {\bibfnamefont {M.~O.}\ \bibnamefont
  {Scully}}\ and\ \bibinfo {author} {\bibfnamefont {M.~S.}\ \bibnamefont
  {Zubairy}},\ }\href@noop {} {\emph {\bibinfo {title} {Quantum Optics}}}\
  (\bibinfo  {publisher} {Cambridge University Press, Cambridge},\ \bibinfo
  {year} {2012})\BibitemShut {NoStop}%
\bibitem [{\citenamefont {MacDonald}\ \emph {et~al.}(2016)\citenamefont
  {MacDonald}, \citenamefont {Hauer}, \citenamefont {Rojas}, \citenamefont
  {Kim}, \citenamefont {Popowich},\ and\ \citenamefont
  {Davis}}]{PhysRevA.93.013836}%
  \BibitemOpen
  \bibfield  {author} {\bibinfo {author} {\bibfnamefont {A.~J.~R.}\
  \bibnamefont {MacDonald}}, \bibinfo {author} {\bibfnamefont {B.~D.}\
  \bibnamefont {Hauer}}, \bibinfo {author} {\bibfnamefont {X.}~\bibnamefont
  {Rojas}}, \bibinfo {author} {\bibfnamefont {P.~H.}\ \bibnamefont {Kim}},
  \bibinfo {author} {\bibfnamefont {G.~G.}\ \bibnamefont {Popowich}}, \ and\
  \bibinfo {author} {\bibfnamefont {J.~P.}\ \bibnamefont {Davis}},\ }\href
  {\doibase 10.1103/PhysRevA.93.013836} {\bibfield  {journal} {\bibinfo
  {journal} {Phys. Rev. A}\ }\textbf {\bibinfo {volume} {93}},\ \bibinfo
  {pages} {013836} (\bibinfo {year} {2016})}\BibitemShut {NoStop}%
\end{thebibliography}%

\end{document}